\documentstyle[epsfig]{aipproc}

\begin{document}
\title{Obscuration model\\ of Variability in AGN}

\author{B. Czerny$^*$, A. Abrassart$^{\dagger}$, S. Collin-Souffrin$^{\dagger}$ and A.-M. Dumont$^{\dagger}$}
\address{$^*$Copernicus Astronomical Center, Bartycka 18, 00-716 Warsaw, Poland\\
$^{\dagger}$DAEC, Observatoire de Meudon, F-92195 France}

\maketitle

\begin{abstract}
There are strong suggestions that the disk-like accretion flow onto massive 
black hole in AGN is disrupted in its innermost part (10-100 Rg), 
possibly due to the 
radiation pressure instability. It may form a hot optically 
thin quasi spherical (ADAF) flow surrounded by or containing denser clouds 
due to the disruption of the disk. Such clouds might be optically thick, with 
a Thompson depth of order of 10 or more.  Within the frame of this cloud 
scenario \cite{collin96,czerny98}, obscuration 
events are expected and the effect would be seen as a variability.
We consider expected random variability due to statistical dispersion 
in location of clouds along the line of sight for a constant covering factor. 
We discuss a simple analytical toy model which provides us with the estimates
of the mean spectral properties and variability amplitude of AGN,
and we support them with radiative transfer computations done
with the use of TITAN code of \cite{dumont99} and NOAR
code of \cite{abrassart99}.
\end{abstract}

\section*{Introduction}

The variability of radio quiet AGN has been established since the early
EXOSAT observations. However, the nature of this variability, observed in the
optical, UV and X-ray band is not clear.

The emission of radiation is caused by accretion of surrounding gas onto a 
central supermassive black hole. The observed variability may be therefore 
directly related to the variable rate of energy
dissipation in the accretion flow. However, it is also possible that the 
observed variability does not represent any significant changes in the
flow. Such an 'illusion of variability' may be created  
if we do not have a full direct view of the nucleus. We explore this 
possibility in some detail. 

Clumpy accretion flow has been suggested by various authors in a physical 
context of gas thermal instabilities or strong magnetic field (e.g. \cite{celotti92,krolik98,torricelli98}).
Here we follow a specific accretion flow pattern described by \cite{collin96}.

We assume that the cold disk flow is disrupted at the distance of 10 - 100
$R_{\rm Schw}$ from the black hole. The resulting clumps of cool material
are large and optically very thick for electron scattering, and they become
isotropically  distributed around the central black hole. Hot plasma 
responsible for hard X-ray emission forms still closer to a black hole, 
perhaps due to cloud collision. We do not discuss the dynamics of cloud
formation but we concentrate on the description of the
radiation produced by such a system. We consider the radiative transfer 
within the clouds and radiative interaction between the clouds and a hot
phase, and we relate the radiation flux and spectra to the variations in
the cloud distribution.

\section*{Variability mechanism} 

Within the frame of the cloud scenario, our line of sight to the hot X-ray
emitting plasma is partially blocked by the surrounding optically thick clouds.
Variations in the cloud distribution lead to two types of phenomena: 
slower variations due to the
systematic change in a total number of clouds and fastest variations due to the random rearrangement of the clouds
without any change of their total number. We concentrate on this second
type of variability.

The amplitude of such a variability is determined by the number of clouds $N$
surrounding a black hole at any given moment and the mean covering factor $C$
of the cloud distribution
$\left({\delta L_{X} \over L_{X}}\right)_{obs} = { C \sqrt {2/N} \over 1 
- C}$.
 
Such variations do not reflect any deep changes in the hot plasma itself so
they are expected to happen without the change in the hard X-ray slope of the
plasma emission. However, the UV variability amplitude caused by the same
mechanism as well as some variations in hard X-rays due to the presence of
the X-ray reflection depend in general on the cloud properties like X-ray
albedo and radiative losses through the unilluminated dark sides
of clouds. We use very detailed radiative transfer in X-ray heated optically
thick clouds in order to estimate those quantities. We develop a complementary
toy model describing the energetics of the entire hot plasma/cloud system
for easy use to estimate the model parameters from the observed variability
amplitudes.

\section*{Radiative transfer solutions for mean spectra}

Two codes are used iteratively in order to compute a mean spectrum emitted
by the clouds distribution. TITAN \cite{dumont99} is designed to solve
the radiative transfer within an optically thick medium, including computations
of the ionization state of the gas and its opacities. NOAR \cite{abrassart99} is a Monte Carlo code designed to follow the hard 
X-ray photons using Monte Carlo method.

\begin{figure}[b!] 
\centerline{\epsfig{file=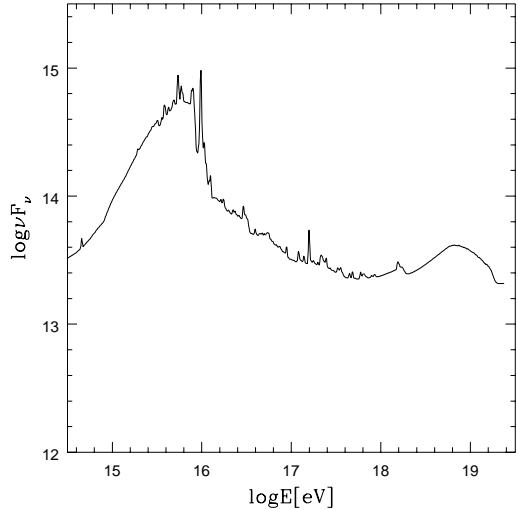,height=3.5in,width=3.5in}}
\vspace{10pt}
\caption{The exemplary spectrum calculated with the coupled codes TITAN and 
NOAR for the following parameters of the shell: number density log$ n =14$, 
column density log $N_H =26$, 
covering factor $C =0.9$. The incident primary radiation was assumed to be
a power law extending from 1 eV to 100 keV, with energy index $\alpha = 1$,
and the ionization parameter $xi=300$. The size of the central source was
neglected.} 
\label{fig1}
\end{figure}

The result of the numerical computation of a single mean spectrum is shown in
Figure \ref{fig1}.
The cloud distribution was assumed to be spherical, with the covering 
factor $C=0.9$, all clouds being located at a single radius. The hot medium in this
computation was 
replaced by a point like source of a primary emission, with flux
normalizations fixed through specification of the ionization parameter $\xi$.
However, the multiple scattering of photons of different clouds was included.

We see that the broad band spectrum clearly
consists of two basic components but there are also detailed spectral features
in UV and soft X-ray band in addition to hard X-ray iron $K_{\alpha}$ line.

\section*{Toy model and variability amplitudes}

In our toy model we replace the radiative transfer computations with analytical
description of the probabilities of the X-ray and UV photon fate. X-ray 
photons can be reflected by
bright sides of the clouds, can escape from the central region towards an 
observer or can be absorbed and provide new UV photons
as well as energy for the emission from the dark sides. UV photon can also
escape, can be reflected and can be upscattered to an X-ray photon by a hot
plasma. All those probabilities are determined by four model parameters:
covering factor $C$, probability of Compton upscattering $\gamma$, X-ray
albedo $a$ and efficiency of dark side emission $\beta_d$. The condition
of compensating for the system energy losses with Compton upscattering relate
those four quantities to the Compton amplification factor. The variability 
amplitude depends also on the number of clouds, $N$.

\begin{figure}[b!] 
\centerline{\epsfig{file=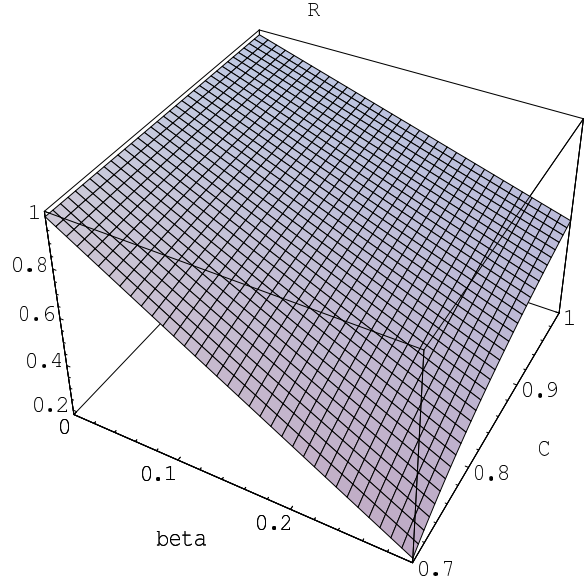,height=3.0in,width=3.0in}}
\vspace{10pt}
\caption{The dependence of the ratio of the normalized variability 
amplitude in $UV$ and 
in X-rays on the dark side loss eficciency $\beta_d$ and covering factor $C$; other parameters: X-ray albedo $a = 0.5$, Compton amplification factor $A = 
4$.} 
\label{fig2}
\end{figure}

Such a model allow us to calculate all basic properties of the stationary
model, like the observed ratio of the X-ray luminosity to the UV luminosity,
the intrinsic ratio of those two quantities as seen by the clouds, the slope of
the hard X-ray emission and the variability amplitudes in UV and X-ray band.

In particular, we can determine the ratio of the normalized variability
amplitudes in X-ray and UV band predicted by our model
$R=\left({\delta L_{UV} \over L_{UV} }\right)_{obs} / \left({\delta 
L_X \over 
L_X}\right)_{obs}$.
In Figure \ref{fig2} we show the dependence of this ratio on the covering
factor $C$ and the efficiency of the dark side energy loss by the clouds.
The dependence on other model parameters was reduced by assuming the X-ray 
albedo $a=0.5$ supported by numerical results and the Compton amplification
factor $A=4$ which well describes the mean hard X-ray spectral slope. 

We see that if the clouds are very opaque  ($\beta_b$ negligible) the normalized
amplitude ratio is always equal 1 within the frame of our model. Significant
dark side energy losses reduce the UV amplitude since they add a constant
contribution to UV flux.

\section*{Discussion}

The presented model well reproduces large observed variability amplitudes
if the covering factor is close to 1. It also explains why large variability
amplitudes are not necessarily accompanied by the change of the slope of 
the hard X-ray emission coming from comptonizing hot plasma. In order to check
whether the model requires unacceptable values of the parameters we confront
the model with the data in the following way.

We apply our toy model to observed variability of four Seyfert 1 galaxies 
extensively monitored in UV and X-ray band (see Table \ref{tab:objects}). The 
variability amplitudes are taken from \cite{goad99,edelson99,nandra98,edelson96,clavel92}, and we estimated the mean X-ray
to UV luminosity ratio as 1/3 in all objects. We assumed the X-ray albedo
$a=0.5$ and the Compton amplification factor as $A=4$. We calculated the
remaining model parameters:  $C$, $N$, $\gamma$, $\beta_d$.  

All four objects are consistent with the model, having relative UV amplitude 
smaller than the relative X-ray amplitude. As expected, the covering factor
(determined by the luminosity ratio) is large and the probability of Compton
upscattering is low either due to small optical depth of the hot plasma or
due to small radial extension of the hot plasma. The required dark side
losses are comparable to the value of 0.20 obtained from the numerical 
solution of the
radiative transfer within a cloud (see Figure \ref{fig1}). Therefore cloud
scenario offers an attractive explanation of the observed variability of AGN
if further observations will confirm that the slope of the direct Compton 
component and its high energy cut-off do not vary.

  \begin{table}
  \caption{Toy model parameters for AGN.
  \label{tab:objects}}
  \begin{tabular}{rrrrrrr}

Object&\multicolumn{1}{c}{$rms_{UV}$} &\multicolumn{1}{c}{$rms_{X}$} & \multicolumn{1}{c}{C} &\multicolumn{1}{c}{ N }&\multicolumn{1}{c}{$\gamma$}&\multicolumn{1}{c}{ $\beta_d$} \\
         &             &        &         &   &    &  \\
   \tableline
   
NGC 3516 & 0.333 & 0.357 &  0.90 &  1180 &  0.047  &  0.04 \\
NGC 7469 & 0.167 & 0.167 &  0.90 &  5400 &  0.044  &  0.00 \\    
NGC 4151 & 0.009 & 0.024 &  0.90 &  2610 &  0.095  &  0.39 \\
NGC 5548 & 0.222 & 0.222 &  0.90 &  3050 &  0.044  &  0.00 \\

  \end{tabular}
  \end{table}

\end{document}